# FDDI: Current Issues and Future Plans[1]

Raj Jain
Digital Equipment Corporation
550 King St. (LKG 1-2/A19)
Littleton, MA 01460
Internet: Jain@Erlang.enet.DEC.Com

ABSTRACT

Key issues in upcoming FDDI standards including low-cost fiber, twisted-pair, SONET mapping, and FDDI follow-on LAN are discussed after a brief introduction to FDDI and FDDI-II.

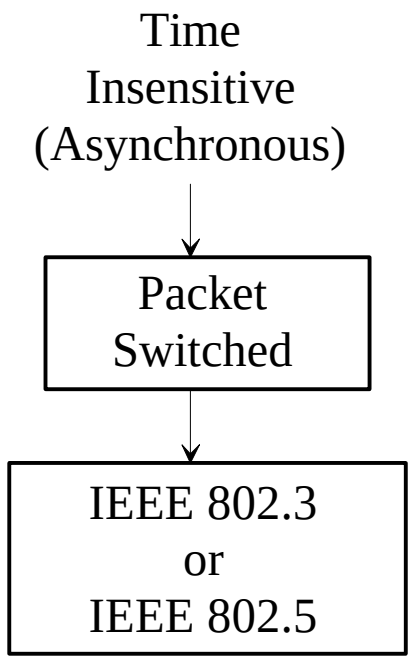

Figure 1: Service provided by IEEE 802.3 and IEEE 802.5 networks.

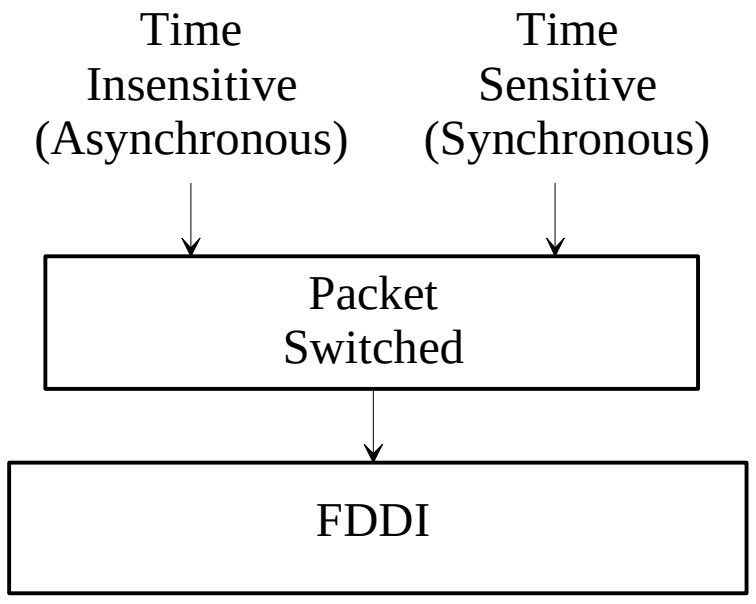

Figure 2: Services provided by FDDI.

## 1 What is FDDI?

Fiber Distributed Data Interface (FDDI) is a set of standards developed by the American National Standards Institute (ANSI) X3T9.5 Task Group. This 100 Mbps local area network (LAN) uses a *timed token* access method to share the medium among stations. The access method is different from the traditional token access method, in that the time taken by the token to walk around the ring is accurately measured by each station and is used to determine the usability of the token.

As shown in Figure 1, older LANs, such as IEEE 802.3/Ethernet and IEEE 802.5/token ring networks, support only asynchronous traffic. FDDI adds synchronous service (see Figure 2). Synchronous traffic consists of delay-sensitive traffic such as voice packets, which need to be transmitted within a certain time interval. The asynchronous traffic consists of the data packets produced by various computer communication applications such as file transfer and mail. These data packets can sustain some reasonable delay and are generally throughput sensitive in the sense that higher throughput (bits or bytes per second) is more important than the time taken by the bits to travel over the network.

An important feature of FDDI, which is also reflected in its name, is its distributed nature. An attempt has been made to make all algorithms distributed in the sense that the control of the rings is not centralized. When any component fails, other components can reorganize and continue to function. This includes fault recovery, clock synchronization, token initialization, and topology control.

In terms of higher layer protocols, FDDI is compatible with IEEE 802 standards such as CSMA/CD (loosely called Ethernet), token rings, and token bus. Thus, applications running on these LANs can be easily made to work over FDDI without any significant changes to upper layer software.

[1] Adapted with permission from Jain (1993).

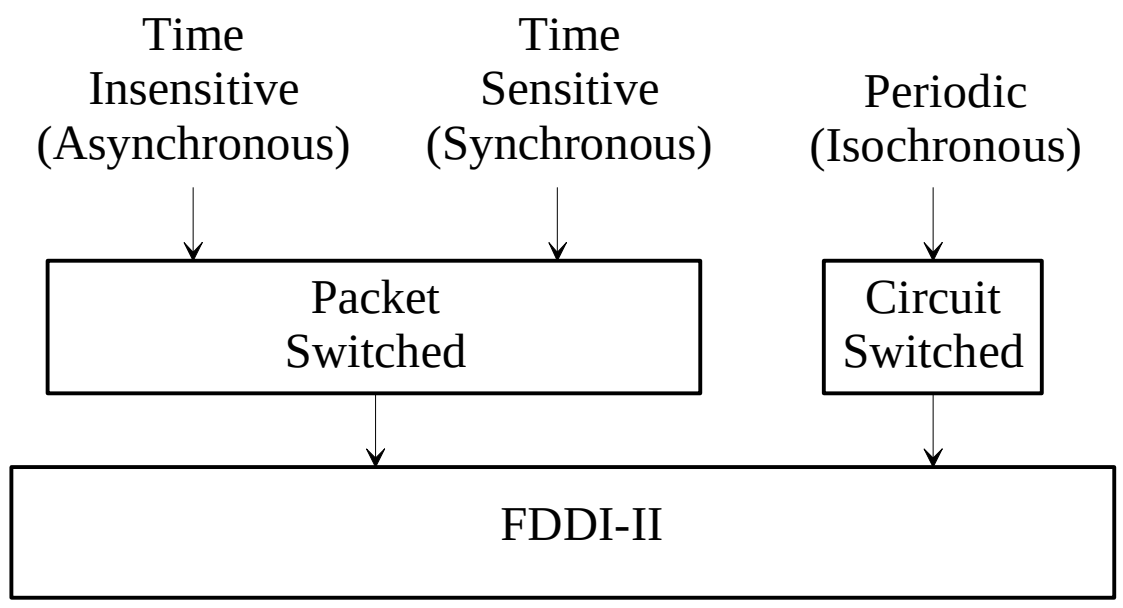

Figure 3: Services provided by FDDI-II.

## 2 FDDI-II

Although the synchronous traffic service provided by FDDI guarantees a bounded delay, the delay can vary. For example, with a target token rotation time (TTRT) value of 165 ms on a ring with 10-$\mu$s latency, a station will get opportunities to transmit the synchronous traffic every 10 $\mu$s at zero load. Under heavy load, occasionally it may have to wait for 330 ms. This type of variation may not suit many constant bit rate (CBR) telecommunication applications that require a strict periodic access. For example, on an ISDN B-channel, which supports one 64-kbps voice conversation, 1 byte is received every 125 $\mu$s. Such circuit-switched traffic cannot be supported on FDDI. If an application needs guaranteed transmission of $n$ bytes every $T$ $\mu$s, or some integral multiples of $T$ $\mu$s, the application is said to require isochronous service.

FDDI-II provides support for isochronous service in addition to asynchronous and synchronous service provided by FDDI as shown in Figure 3.

Like FDDI, FDDI-II runs at 100 Mbps. FDDI-II nodes can run in FDDI or basic mode. If all stations on the ring are FDDI-II nodes, then the ring can switch to the hybrid mode in which isochronous service is provided in addition to basic mode services. However, if there is even one station on the ring that is not an FDDI-II node, the ring cannot switch to the hybrid mode and will keep running in the basic mode. In the basic mode on FDDI-II, synchronous and asynchronous traffic is transmitted in a manner identical to that on FDDI. Isochronous service is not available in the basic mode.

Most multimedia applications such as video conferencing, real-time video, and entertainment video can be supported on FDDI since the required time guarantee is a few tens of milliseconds. This can be easily guaranteed with the synchronous service and a small TTRT. Since TTRT cannot be less than the ring latency, applications requiring time bounds less than twice the ring latency cannot be supported by FDDI. Similarly, applications requiring strict periodic access will require FDDI-II. The main problem facing FDDI users is that even if only one or two stations require isochronous service, hardware on all stations on the ring would have to be upgraded to FDDI-II.

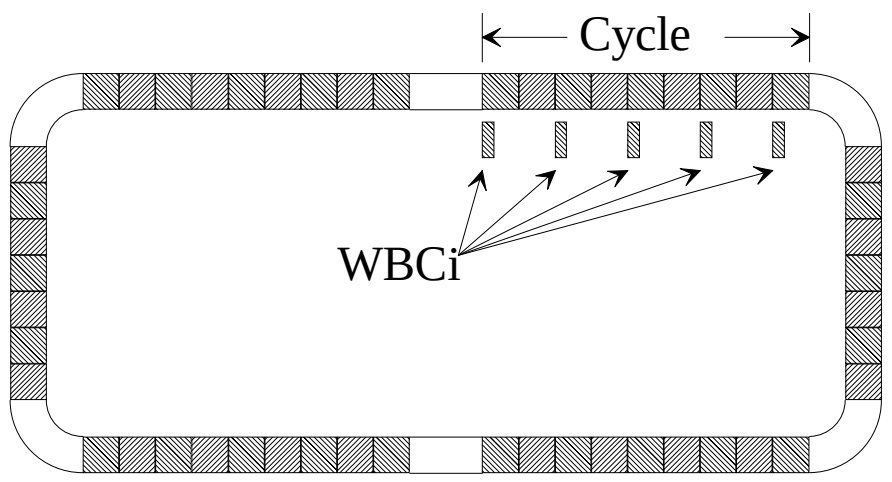

Figure 4: Cycles.

To service periodic isochronous requests, FDDI-II uses a periodic transmission policy in which transmission opportunities are repeated every 125 $\mu$s. This interval has been chosen because it matches the basic system reference frequency clock used in public telecommunications networks in North America and Europe. At this interval, a special frame called a *cycle* is generated. At 100 Mbps, $1562\frac{1}{2}$ bytes can be transmitted in 125 $\mu$s. Of these, 1560 bytes are used for the cycle and $2\frac{1}{2}$ bytes are used as the intercycle gap or cycle preamble. At any instant, the ring may contain several cycles as shown in Figure 4.

The bytes of the cycles are preallocated to various channels (for communication between two or more stations) on the ring. For example, a channel may have the right to use the 26th and 122nd bytes of every cycle. These bytes are reserved for the channel in the sense that if the stations owning that channel do not use it, other stations cannot use it and the bytes will be left unused.

The 1560 bytes of the cycle are divided into 16 *wideband channels* of 96 bytes each. Each wideband channel (WBC) provides a bandwidth of 96 bytes per 125 $\mu$s or 6.144 Mbps. This is sufficient to support one television broadcast, four high-quality stereo programs, or ninty-six telephone conversations.

Some of the 16 wideband channels may be allocated for packet mode transmissions and the others for isochronous mode transmissions. For example, channels 1, 5, and 7 may be used for packet mode transmissions or packet switching, and channels 2, 3, 4, 6, and 8 through 15 may be used for isochronous mode transmissions or circuit switching. It is possible to allocate all wideband channels for circuit switching alone or packet switching alone. The allocation is made using station management protocols, which have not yet been defined.

## 3 Low-Cost Fiber

After the initial FDDI specifications were completed in 1990, it was realized that one of the impediments to rapid deployment of FDDI was the high cost of optical components. To switch from the lower speed technology of Ethernet or token ring, it was necessary to rewire the building, install FDDI concentrators, install FDDI adapters in systems, install new software, and so on. Although the cost of all components is continuously decreasing, it is still high. Therefore, a standard effort has begun to find a low-cost alternative.

This effort has resulted in a new *media-dependent physical layer* (*PMD*) standard called *Low Cost Fiber PMD* (*LCF-PMD*). As the name suggests, originally the committee intended to find a fiber that was cheaper than the 62.5/125 multimode fiber used in the standard FDDI. A number of alternatives such as plastic fiber and 200/230 $\mu$m fiber were considered but were quickly rejected when it was realized that the real expense was in the devices (transmitters and receivers) and not in the fiber. A search for lower powered devices then began.

LCF-PMD allows low-cost transmitter and receiver devices to be used on any FDDI link. These devices are cheaper because they have more relaxed noise margins and are either lower powered or less sensitive than those specified in the original PMD (which we prefer to call MMF-PMD, MMF stands for multimode fiber). The specification has been designed for links up to 500 m long (compared to 2 km in MMF-PMD). This distance is sufficient for most intrabuilding applications.

Only the interbuilding links that are longer than 500 m need to pay the higher cost of MMF-PMD devices. Any combination of LCF, MMF, single-mode fiber (SMF), SONET, and copper links can be intermixed in a single FDDI network as long as the distance limitations of each are carefully followed.

Table 1 provides a comparison of the key design decisions for LCF and MMF PMDs. These are explained further below.

1. *Wavelength*: LCF uses the 1300-nm wavelength, which is the same as in multimode and single-mode PMDs. Initially, an 850-nm wavelength was suggested because 850-nm devices are used in fiber optic Ethernet (IEEE 802.3 10BASE-F) and token ring (IEEE 802.5J) networks. They are sold in large volume and so are much cheaper than 1300-nm devices. However, this would have introduced a problem of incompatibility because the users would have to remember (and label) the source wavelength and use the same wavelength device at the receiving end. This would also have caused the receiver to be replaced every time the transmitter was replaced. With 1300 nm at both ends, the user need only worry about the distance. As long as the distance is less than 500 m, the two ends can use any combination of LCF and MMF devices.

2. *Fiber*: LCF specifies 62.5/125-$\mu$m graded-index multimode fiber — the same as that specified in MMF-PMD. Initially plastic fibers and 200/230-$\mu$m step-index fibers were considered. Plastic fibers are inexpensive but they have a high attenuation. Using plastic fibers would have severely limited the distance. Two-hundred micron fibers have a larger core, which allows for a larger amount of power to be coupled in the fiber. The connectors and splices for these fibers are also cheaper since no active alignment is required. However, the large diameter of the core implies more dispersion and therefore lower bandwidth. For 200-$\mu$m fiber, a bandwidth-distance product of 30 MHz-km was predicted while 80 MHz-km (800 MHz over 100 m) has been measured. Connectors for 200-$\mu$m fiber are 30 percent cheaper; transceivers are 70 percent cheaper even while producing 10 times more power than the standard FDDI. The larger power is required because the 200-$\mu$m fiber has an attenuation of 16 dB/km compared to 2 dB/km for 62.5/125-$\mu$m fiber.

   The main problem with 200/230-$\mu$m fibers is that intermixing them with 62.5/125 fibers on the same link causes a significant amount of power loss. When it was realized that a 50 percent cost reduction goal could be achieved by simply changing the transmitting and receiving power levels by 2 dBm, all efforts to change the fiber came to a halt.

3. *Connector*: The duplex connector specified in MMF-PMD was designed specifically for FDDI. Due to its low-volume production, its cost is high. Significant savings can be obtained by using other simplex connectors. In fact, many FDDI installations already use the simplex-ST connector. The LCF committee wanted to use a duplex connector to avoid the problem of misconnections (resulting in two transmitters being connected to each other). A duplex-SC connector, shown in Figure 5, was proposed. SC, which stands for subscriber connector, is a Japanese standard. It is an augmentation of the FC connector. The SC connector was developed in 1984 to provide a push-pull interface, which reduces the space required between the connectors (compared to the case if the connector has to be rotated by fingers). As a result, a large number of connectors can be placed side by side. SC has a connector loss of 0.3 dB and a return loss (reflection) of −43 dB. In the US, ST connectors are more popular than SC or FC connectors. A number of companies have proposed duplex-ST connector designs with specifications matching those of the duplex-SC. After much heated debate, termed "Connector War II," duplex-SC was voted the main selection, with duplex-ST being the recommended alternate.

Table 1: Low-Cost Fiber versus Multimode Fiber PMD

| Issue | MMF | LCF |
| --- | --- | --- |
| Wavelength | 1300 nm | 1300 nm |
| Fiber | 62.5/125 multimode | 62.5/125 multimode |
| Transmitter power | Max −20 dBm | Max −22 dBm |
| Receiver power | Min −31 dBm | Min −29 dBm |
| Connector | Duplex-FDDI | Duplex-SC or duplex-ST |
| Connector keying | Port and polarity | No port. Polarity only. |

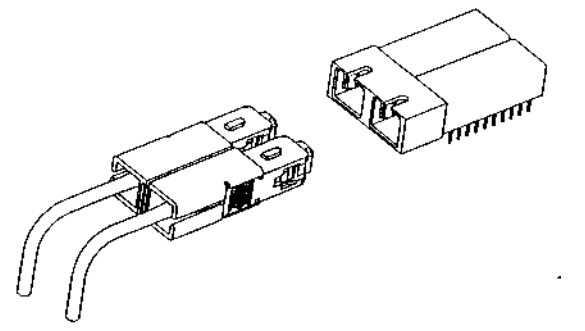

Figure 5: Duplex-SC connector

4. *Transmitter/Receivers*: Reducing the transmitted power and the dynamic range (even slightly) reduces the cost significantly. LCF-PMD reduces the required transmit power by 2 dBm and the receiver dynamic range by 2 dBm. The transmitted power range is (−22, −14) dBm while the received power range is (−29, −14) dBm. This means that the maximum loss allowed in the fiber is only 7 dB (=29−22) instead of 11 dB. This is sufficient for a 500 m link.

5. Rise/Fall Times: Since LCF-PMD uses the same fiber as MMF-PMD but the link length has been decreased from 2 km to 500 m, the pulse broadening caused by fiber dispersion is less. The change then in pulse rise and fall times due to the fiber is not as much. The time thus saved has been allocated to transmitters and receivers to reduce their cost. Thus, LCF transmitters are allowed to have a rise/fall time of 4.0 ns compared to 3.5 ns for MMF transmitters. Thus, lower quality (hence, cheaper) transmitters can be used. Similarly, LCF receivers are required to receive pulses with a rise/fall time below 4.5 ns compared to 5 ns for MMF receivers. This again means less work (hence, lower cost) for the receivers.

## 4 Twisted-Pair PMD (TP-PMD)

As soon as the initial FDDI products started appearing on the market, the realization was made that one impediment to FDDI acceptance was that it required users to rewire their buildings with fibers. Even if you only need to connect two nearby pieces of equipment on the same floor, you will need to install fibers. Rewiring a building is a major expense and is not easy to justify unless the technology is well proven or absolutely necessary.

Besides the wiring expense, the optical components used in FDDI equipment are also very expensive compared to the electronic components used in other existing LANs. This led several manufacturers to look into the possibility of providing 100-Mbps communication on existing copper wiring. It was determined that 100 Mbps transmission using high-quality (shielded or coax) copper cables is feasible at a much lower cost than that of the fiber, particularly if the distance between nodes is limited to 100 m. Other manufacturers later found ways to transmit 100 Mbps on unshielded twisted pair (UTP), which is used in telephone wiring, up to 50 m.

An FDDI ring can have a mixture of copper and fiber links. Therefore, short links used in office areas can use existing copper wiring installed for telephones or other LAN applications. This results in considerable cost savings and quicker migration from lower speed LANs to FDDI. Proprietary coaxial cable and shielded twisted-pair (STP) products, which support FDDI links of up to 100 m, are already available. Over 98 percent of the data cable running in offices is less than 100 m and 95 percent is less than 50 m. These can be easily upgraded to run at 100 Mbps.

FDDI twisted-pair PMD is still under development. The major design issues are:

1. *Categories of Cables*: Sending a 125-Mbps signal over a coaxial cable or STP is not as challenging as on UTP. Given the preponderance of UTP cabling to the desktop in most offices, it is clear that allowing FDDI on UTP, however difficult, will be a major win for FDDI. The first issue was whether we should have different coding methods for UTP and STP or one standard covering both. A decision has been made to have one standard for both. Which categories of UTP should it cover is the next issue. While it is easier to handle data-grade twisted pair (EIA Category 5), allowing Category 3 cable would introduce more complexity.

2. *Power Level*: The attenuation (loss) of signal over copper wires increases at high frequency. To maintain a high signal-to-noise ratio, one must either increase the signal level (more power) or use special coding methods to produce lower frequency signals. Increased power results in increased interference and therefore special coding methods are required.

3. *Electromagnetic Interference*: The main problem caused by high-frequency signals over copper wires is the electromagnetic interference (EMI). After 4b/5b encoding, the FDDI signal has a bit rate of 125 Mbps. With NRZI encoding this results in a signal frequency of 62.5 MHz. At this frequency range, the copper wire acts as a broadcasting antenna. The electromagnetic radiations from the wire interfere with radio and television transmissions. The interference increases with the signal level. Federal Communications Commision (FCC) places strict limits on such electromagnetic interference (EMI). This severely limits the power that the FDDI transmitters can use, which in turn means that the distance at which the signal becomes unintelligible is also limited.

   One solution to the EMI problem is to use shielded wire (STP) or coaxial cable. These wires have a special metallic shield surrounding the wires that prevents interference. Another solution is to use special coding techniques that result in a lower frequency signal. The advantage of this second approach is that the unshielded twisted-pair wires, which reach all desks, can be used for FDDI. The issue of coding has now been resolved and a three-level coding called multilevel transmission 3 (MLT-3) has been selected. This reduces the signal frequency by a facotr of 2.

4. *Scrambling*: Even though MLT-3 (and other) encoding schemes reduce the signal frequency, they are not sufficient to meet the FCC EMI requirements for UTP. One way to reduce interference is to scramble the signal so that the energy is not concentrated at one frequency. Instead, it is distributed uniformly over a range of frequencies.

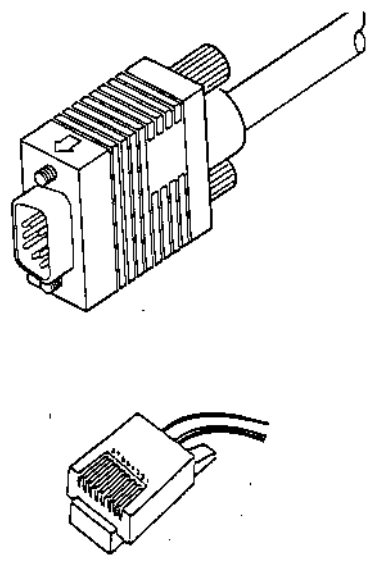

Figure 6: Connectors proposed for TP PMD

5. *Connectors*: Figure 6 shows the RJ-45 and DB-9 connectors proposed for TP-PMD. Both are popular connectors. They are available at a very low price due to their widespread use in computer and communication industries.

# 5 FDDI on SONET

"SONET" stands for Synchronous Optical Network. It is a standard developed by ANSI and Exchange Carriers Standards Association (ECSA) for digital optical transmission. If you want to lease a fiber-optic line from your telephone company, it is likely to offer you a "SONET link" instead of a dark fiber link. A SONET link allows the telephone company to divide the enormous bandwidth of a dark fiber among many of its customers. Thus, a SONET link is much cheaper compared to a dark fiber link.

The SONET standard has also been adopted by CCITT. There are slight differences between the CCITT and ANSI versions. The CCITT version is called Synchronous Digital Hierarchy (SDH).

A SONET system can run at a number of predesignated data rates. These rates are specified as STS-N rates in the ANSI standard. STS-N stands for Synchronous Transport Signal level N. The lowest rate STS-1 is 51.84 Mbps. Other rates of STS-N are simply N times this rate. For example, STS-3 is 155.52 Mbps and STS-9 is 466.56 Mbps. Table 2 lists the complete hierarchy. The corresponding rate at the optical level is called Optical Carrier level N (OC-N). Since each bit results in one optical pulse in SONET (no 4b/5b type of coding is used), the OC-N rates are identical to STS-N rates.

Table 2: SONET/SDH Signal Hierarchy

| ANSI Designation | Optical Signal | CCITT Designation | Data Rate (Mbps) | Payload Rate (Mbps) |
|---|---|---|---|---|
| STS-1 | OC-1 | | 51.84 | 50.112 |
| STS-3 | OC-3 | STM-1 | 155.52 | 150.336 |
| STS-9 | OC-9 | STM-3 | 466.56 | 451.008 |
| STS-12 | OC-12 | STM-4 | 622.08 | 601.344 |
| STS-18 | OC-18 | STM-6 | 933.12 | 902.016 |
| STS-24 | OC-24 | STM-8 | 1244.16 | 1202.688 |
| STS-36 | OC-36 | STM-12 | 1866.24 | 1804.032 |
| STS-48 | OC-48 | STM-16 | 2488.32 | 2405.376 |
| STS-96 | OC-96 | STM-32 | 4976.64 | 4810.176 |
| STS-192 | OC-192 | STM-64 | 9953.28 | 9620.928 |

For the CCITT/SDH standard, the data rates are designated STM-N (Synchronous Transport Module level N). The lowest rate STM-1 is 155.52 Mbps. Other rates are simply multiples of STM-1.

In both cases, some bandwidth is used for network overhead. The data rate available to the user, called the payload rate, is also shown in Table 2.

SONET physical-layer mapping (SPM) takes the output of the current FDDI physical layer, which is a 4b/5b encoded bit stream, and places it in appropriate bits of an STM-1 synchronous payload envelope (SPE). An STM-1 SPE consists of 2349 bytes (arranged as 9 rows of 261 bytes each). Of these, 9 bytes are used for path overhead. Since one SPE is transmitted every 125 $\mu$s, the available bandwidth is $(2349\times 8)/125$ or 139.264 Mbps. This is more than the 125 Mbps required for FDDI. The extra bits are used for network control purposes and as stuff bits for overcoming clock jitter.

SONET uses a simple NRZ encoding of bits. In this coding, a 1 is represented as high level (light on) and a 0 is represented as low level (light off). One problem with this coding is that if too many 1's (or 0's) are transmitted, the signal remains at on (or off) for a long time, resulting in a loss of bit clocking information. To solve this problem, the SONET standard requires that all bytes in a SONET signal be scrambled by a frame synchronous scrambler sequence of length 127 generated by the polynomial $1+x^6+x^7$. Certain overhead bytes

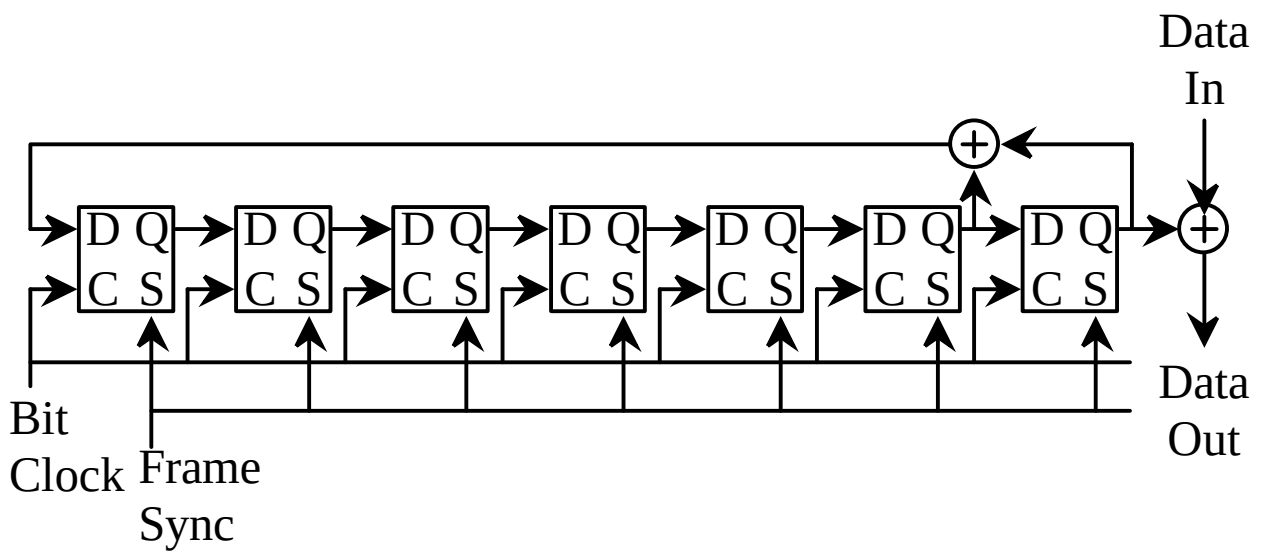

Figure 7: Shift-register implementation of a SONET scrambler.

are exempt from this requirement.

The scrambler consists of a sequence of seven shift registers as shown in Figure 7. At the beginning of a frame, a seed value of 1111111 is loaded in the register. As successive bits arrive, the contents of shift registers are shifted and the sixth and seventh registers' contents (this corresponds to terms $x^6$ and $x^7$, respectively) are exclusive-or'ed and fed back to the first register (corresponding to the first term in the polynomial). The output of the final shift register is a random binary pattern, which is exclusive-or'ed to the incoming information bits.

The scrambling operation is equivalent to exclusive-oring of the bits with a particular 127-bit sequence. The sequence is highly random and does not contain long sequences of 1's or 0's. Therefore, this is expected to increase the frequency of transitions in the resulting stream. However, if the user data pattern is identical to any subset of this sequence, the resulting stream will have all 1's in the corresponding bit positions. Similarly, if the user data pattern is an exact complement of any subset of this sequence, the resulting stream will have all 0's in the corresponding bit positions.

One of the key issues in the design of the FDDI-to-SONET mapping was to ensure that the FDDI signal pattern does not result in long series of 1's or 0's after scrambling. Two steps have been taken for this purpose. First, several fixed stuff bits are used throughout the SPE to break up the FDDI stream. As a result, FDDI data cannot affect more than 17 contiguous bytes. Even the 17-byte string has one bit that is a stuff control bit; therefore, not under user control. Second, the scrambler sequence was analyzed to find the longest possible valid 4b/5b pattern that will match (or complement) a portion of the scrambler sequence. The longest possible match for random sequences of FDDI data or control symbols and the SONET scrambler sequence is 58 bits (7.25 bytes) of valid symbols. Thus, it is not possible for an FDDI user to cause serious errors in the SONET network by simply sending a data pattern.

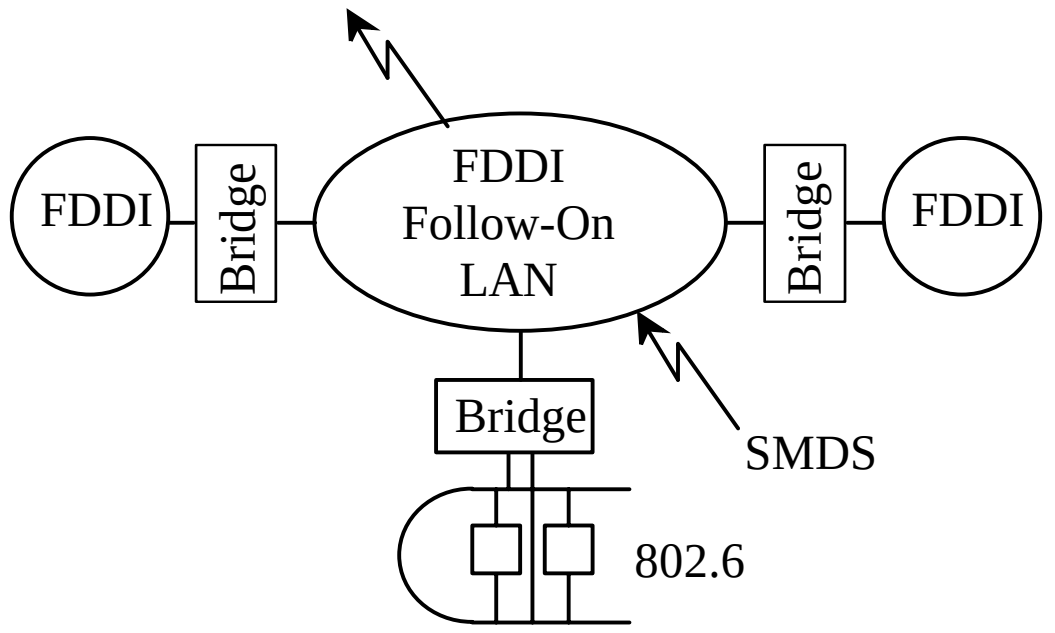

Figure 8: FDDI Follow-On network as a backbone.

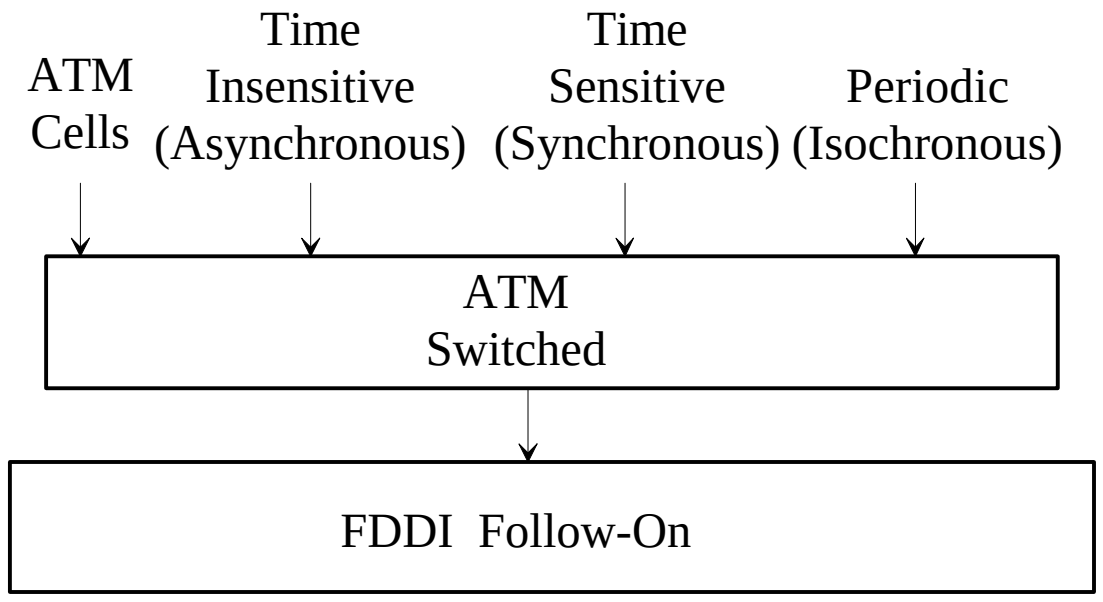

Figure 9: Services provided by the FDDI Follow-On LAN

## 6 FDDI Follow-On LAN

Both FDDI and FDDI-II run at 100 Mbps. To connect multiple FDDI networks, there is a need for a higher speed backbone network. The FDDI standards committee has realized this need and has started working on the design of the next generation of high-speed networks. The project is called FDDI Follow-On LAN (FFOL). Currently, the project is in its infancy and not much has been decided. All of the information presented here is preliminary and subject to rapid change.

The key goal of FFOL is to serve as a backbone network for multiple FDDI and FDDI-II networks. This implies that it should provide at least the packet switching and circuit switching services provided by FDDI-II. For a backbone network to be successful, it should be able to carry the traffic on a wide variety of networks. Other networks that run at speeds close to that of FDDI and that are expected to use FFOL are broadband integrated services digital networks (B-ISDN), which use asynchronous transfer mode (ATM). ATM networks use small fixed size cells. FFOL is expected to provide an ATM service that will allow the cells to be switched among ATM networks. This will allow IEEE 802.6 dual queue dual bus (DQDB) networks also to use FFOL as the backbone (see Figure 8). Easy connection to B-ISDN networks is one of the key goals of FFOL.

The key issues in the design of a high-speed network are:

1. *Data Rate*: By the time FFOL is ready, multimode fibers are expected to be in common use because of FDDI. It is desirable that users be able to use the installed fiber in FFOL. It is well known from the FDDI design that these multimode fibers have the capacity to run 100 Mbps (125 Mbps signaling rate) up to 2 km. Therefore, they can also carry a signal of 1.25 Gbps up to 200 m or 2.5 Gbps up to 100 m. The latter (100 m) covers the length of the horizontal wiring supported by ANSI/EIA/TIA 568 for commercial building wiring standards. Limiting FFOL to below 2.5 Gbps will allow much of the installed multimode fiber in the buildings to be switched from FDDI to FFOL.

   To carry telecommunication network traffic, FFOL should support data rates that are compatible with SONET. FFOL will be designed to be able to efficiently exchange traffic at STS-3 (155.52 Mbps), STS-12 (622.08 Mbps), STS-24 (1.24416 Gbps), and STS-48 (2.48832 Gbps).

2. *Media Access Modes*: The term *media access modes* refers to the traffic switching modes supported by a network. FDDI supports three different modes of packet-switching: synchronous, asynchronous, and restricted asynchronous. Depending upon the delay and throughput requirements, an application can choose any one of these three media access modes. FDDI-II adds support for periodic (isochronous) traffic that normally requires circuit switching. FFOL is expected to support these modes. In addition, as shown in Figure 9, it is expected to explicitly support ATM switching as well. ATM switching is slightly different from packet switching. All ATM cells are the same size, the switching instants are fixed, and a slotted network design is generally used. One proposal calls for using cycles similar to those in FDDI-II and allocating some wideband channels for ATM. Another alternative is to use an ATM base to support isochronous, ATM, and packet switching.

3. *Media Access Method*: Media access method refers to the rules for sharing the medium. Token, timed token, and slotted access are examples of the media access methods used in IEEE 802.5 token ring, FDDI, and Cambridge ring, respectively. The token access method is not useful for long distances. Its deficiency can be easily seen by considering what happens at zero load or at very high loads. Even when nothing is being transmitted, a token must be captured. This may take as long as the round trip delay around the network (ring latency). If the ring latency is $D$, the average access time at zero load is $D/2$. For networks covering large distances, this may be unacceptable. At high load, the maximum relative throughput (or efficiency) of the timed token access method is:

$$\text{Efficiency} = \frac{n(T-D)}{nT+D}$$

where, $T$ is the target token rotation time. The efficiency decreases as the ring latency $D$ increases. In the extreme case of $D=\infty$, the efficiency is zero. Access methods, whose efficiency reduces with propagation delay, are also sensitive to network bit rate. Their efficiency decreases as the bit rate increases. Since the geographic extent covered by a backbone FFOL network is expected to be large, FFOL is expected to select a media access method that is relatively insensitive to the propagation delay and network bit rate.

4. *Physical Encoding*: FDDI uses a 4b/5b encoding, which allows 4 data bits to be combined into one symbol. The electronic processing is done either on symbols or on symbol pairs. These are known as symbol-wide and byte-wide implementations, respectively. Assuming a symbol-wide implementation, the electronic circuits run at 25 Mbps for 100 Mbps FDDI. At one Gbps, using the same encoding, the electronic devices will have to run at 250 Mbps. Such devices are expensive. Using larger symbol sizes such as 8/10, 16/20, or 32/40 allows parallel processing using low-speed electronic circuits. Large symbol size also allows more control symbols. These control symbols are useful for framing, fault recovery, and physical connection management.

5. *Frame Stripping Method*: Destination stripping allows spatial reuse such that the space on the media freed by the destination can be used by the destination or other succeeding nodes. Several simultaneous transmissions can be in progress in networks implementing spatial reuse. Thus, the total network throughput can be as much as $n$ times the network bandwidth, where $n$ is the number of simultaneous transmissions.

   Destination stripping is generally used in non-token rings such as register insertion rings and slotted rings. In networks using simple token access methods, multiple simultaneous transmission is not possible since each transmitting station needs a token and there is only one token. Token networks, therefore, use source stripping.

   Proposals have been made for FFOL to use destination stripping. This is because the extent of the network will be large and tying up the whole medium for one transmission is not desirable.

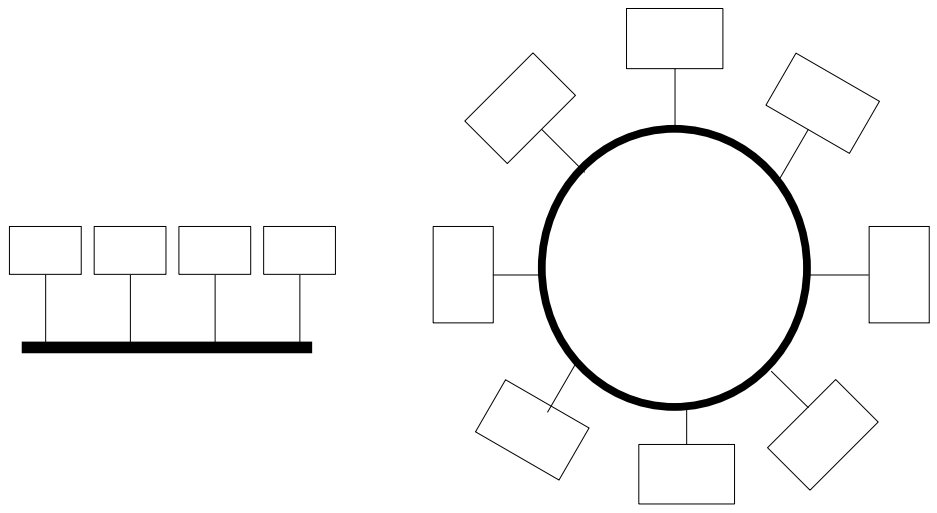

Figure 10: Shared-media distributed-switching approach.

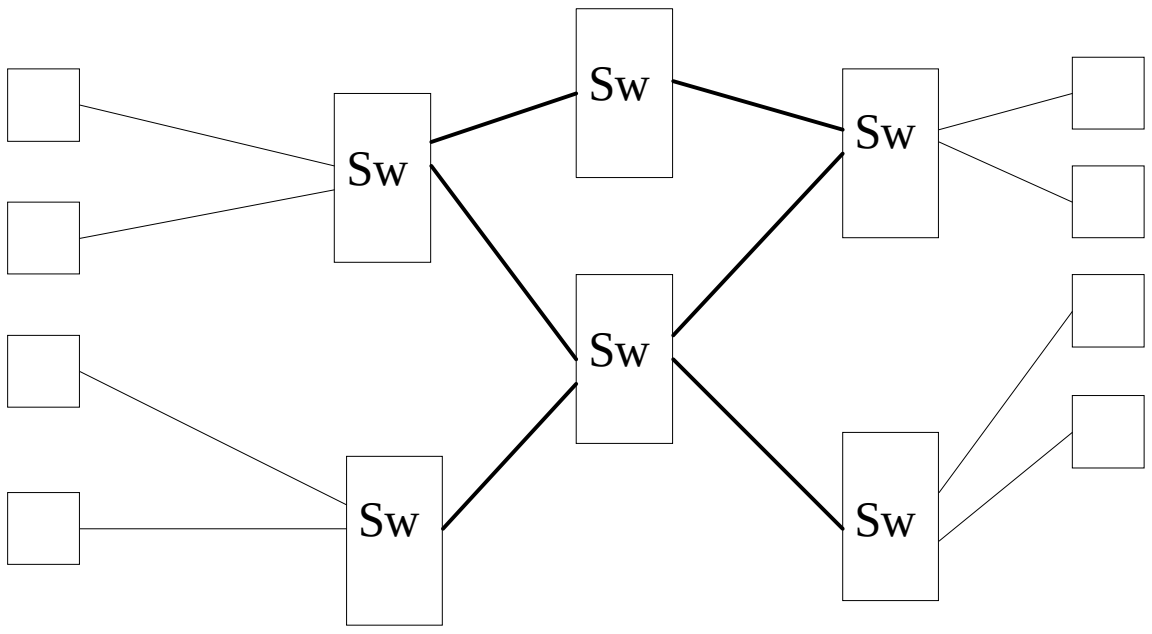

Figure 11: Shared-switching distributed-media approach.

6. *Topology*: FFOL is expected to allow the dual ring of trees physical topology that is supported by FDDI. Additional topologies may be allowed. Segments of public networks may be included in the FFOL networks. In current FDDI, only SONET links are allowed.

   All LANs are designed so that the responsibility for ensuring that the packet is delivered to the correct destination is shared by all nodes. In this case, as shown in Figure 10, the switching is distributed and the medium is shared. Another alternative, shown in Figure 11, is to distribute the medium and share switches.

   The advantage of this latter approach is that not all end stations need to pay the cost of a high-speed connection. Their links can be upgraded to higher speeds only when necessary. The end systems are simple and most of the design complexity is in the switches. There can be several parallel transmissions at all times. Thus, the total throughput of the network is several times the bandwidth of any one link. For example, it is possible to get a total network throughput of several Gbps with all links having a bandwidth of only 100 Mbps. Notice that most the telecommunication networks and wide-area computer networks use the switch-based approach. Even in high-speed LANs, there is a trend towards a switch-based mesh topology. It is not clear whether FFOL will consider a mesh topology.

## 7 Summary

FDDI is the next generation of high-speed networks. It is an ANSI standard that is being adopted by ISO and is being implemented all over the world. It allows communication at 100 Mbps among 500 stations distributed over a total cable distance of 100 km.

FDDI will satisfy the needs of organizations needing a higher bandwidth, a larger distance between stations, or a network spanning a greater distance than the Ethernet or IEEE 802.5 token-ring networks. It provides high reliability, high security, and noise immunity. It supports data as well as voice and video traffic.

FDDI-II provides all services provided by FDDI but adds support for isochronous traffic.

Low-cost fiber PMD allows cheaper fiber links not by using cheaper fiber but by using low-powered transceivers. The net link budget has been reduced from 11 dB to 7 dB. The reduced power allows such links to be used only if the link length is less than 500 m. Most intrabuilding links are within this distance range. The cost of the connector has also been reduced by selecting duplex versions of popular simplex connectors. These connectors are required to only have polarity keying so that an untrained user cannot misconnect a transmitter to another transmitter. No port type keying is required.

Stallings (1992) has a chapter devoted to the SONET standard. The analysis of the SONET scrambler for FDDI mapping is presented in Rigsbee (1990).

Standardization of FDDI on copper will reduce its cost considerably and help bring FDDI to the desktop.

The next higher speed version of FDDI, called FDDI Follow-On LAN, and running at 600 Mbps to 1.2 Gbps speed, is currently being discussed.

## 8 Further Reading

The FDDI protocols are described in a number of ANSI standards and working documents. These standards are also being adopted as ISO standards.

Much of this article has been excerpted from Jain (1993).

See Burr and Ross (1984), Ross and Moulton (1984), Ross (1986, 1989, and 1991), and Hawe, Graham, and Hayden (1991) for an overview of FDDI.

Caves and Flatman (1986), Teener and Gvozdanovic (1989), and Ross (1991) provide an overview of FDDI-II.

Ginzburg, Mallard, and Newman (1990) discuss some of the problems in transmitting high bandwidth signal over copper.

FFOL requirements and design considerations are summarized in Ocheltree, Horvath, and Mityko (1990) and in Ross and Fink (1992).

## 9 Acknowledgement

Thanks to Bill Cronin and Paul Koning for useful feedback on an earlier version of this paper.

**Side Bar:** *Is FDDI a Misnomer?*

ANSI Task Group X3T9.5 was formed in 1979 to provide a high-performance I/O channel called Local Distributed Data Interface (LDDI). The idea of using optical fibers was first raised in subcommittee X3T9.5 at the October 1982 meeting and subsequently LDDI standard was abandoned and a new effort based on fiber was begun. This new standard was named Fiber Distributed Data Interface or FDDI.

Initially the standard was expected to be used only on a fiber medium in a fully distributed manner for data transmission. It was expected to only specify an interface similar to SCSI (Small Computer System Interface). The features of FDDI have slowly been extended to meet diverse needs and now the name FDDI has actually become a misnomer. A more appropriate name for the current FDDI standards would be **** (four asterisks), where each asterisk stands for a wild-card in that position. FDDI is now an any-media centralized-or-distribued, any-traffic (voice, video, or data) LAN-or-MAN-or-interface.

FDDI standards now cover non-fiber media including copper wires. Since FDDI-II uses a centralized ring master station, it is not fully distributed peer-to-peer protocol. Data was never considered to be the only traffic on FDDI. Even initial versions have features for voice, video, and other telephony applications. Finally, FDDI is a full featured network and not just another bus interface.